\documentclass[aps,prd,amsmath,amssymb,superscriptaddress,preprintnumbers,twocolumn,nofootinbib,10pt]{revtex4-1}
\pdfoutput=1

\usepackage{graphicx}
\usepackage{dcolumn}
\usepackage{bm}
\usepackage{booktabs}
\usepackage{amsmath}
\usepackage{amssymb}
\usepackage{multirow}
\usepackage{url}
\usepackage{threeparttable}
\usepackage{array}
\usepackage{adjustbox}
\usepackage{makecell}
\usepackage{tabularx}
\usepackage[normalem]{ulem}
\usepackage{xcolor}
\usepackage[bottom]{footmisc}
\usepackage[colorlinks=true, linkcolor=blue, citecolor=blue]{hyperref}
\usepackage{float}

\newcolumntype{C}{>{\centering\arraybackslash}X}

\begin{document}

\title{Neutrino mass constraints in the Schwarzschild-de Sitter black-hole dark energy model with ACT DR6 and DESI DR2 data}

\author{Sheng-Han Zhou}
\affiliation{Liaoning Key Laboratory of Cosmology and Astrophysics, College of Sciences, Northeastern University, Shenyang 110819, China}

\author{Tian-Nuo Li}\thanks{Corresponding author}\email{litiannuo@stumail.neu.edu.cn}
\affiliation{Liaoning Key Laboratory of Cosmology and Astrophysics, College of Sciences, Northeastern University, Shenyang 110819, China}

\author{Guo-Hong Du}
\affiliation{Liaoning Key Laboratory of Cosmology and Astrophysics, College of Sciences, Northeastern University, Shenyang 110819, China}

\author{Yi-Min Zhang}
\affiliation{Liaoning Key Laboratory of Cosmology and Astrophysics, College of Sciences, Northeastern University, Shenyang 110819, China}

\author{Zhao-Yu Li}
\affiliation{Liaoning Key Laboratory of Cosmology and Astrophysics, College of Sciences, Northeastern University, Shenyang 110819, China}

\author{\\Jing-Fei Zhang}
\affiliation{Liaoning Key Laboratory of Cosmology and Astrophysics, College of Sciences, Northeastern University, Shenyang 110819, China}

\author{Xin Zhang}\thanks{Corresponding author}\email{zhangxin@mail.neu.edu.cn}
\affiliation{Liaoning Key Laboratory of Cosmology and Astrophysics, College of Sciences, Northeastern University, Shenyang 110819, China}
\affiliation{MOE Key Laboratory of Data Analytics and Optimization for Smart Industry, Northeastern University, Shenyang 110819, China}
\affiliation{National Frontiers Science Center for Industrial Intelligence and Systems Optimization, Northeastern University, Shenyang 110819, China}

\begin{abstract}

Recent DESI observations have posed new challenges to $\Lambda$CDM, showing a preference for dynamical dark energy and yielding neutrino mass constraints within $\Lambda$CDM that approach the lower bound allowed by neutrino oscillation experiments. In this work, we investigate cosmological constraints on the key neutrino parameters, $\sum m_\nu$ and $N_{\rm eff}$, within the Schwarzschild--de Sitter black-hole dark energy (SdSDE) framework. We use cosmic microwave background (CMB) data from Planck and ACT DR6, baryon acoustic oscillation data from DESI DR2, and type Ia supernova data from DES-Dovekie and PantheonPlus. We find that SdSDE scenarios prefer a positive neutrino mass whenever $\sum m_\nu$ is allowed to vary. Using CMB+DESI+DES-Dovekie data, we obtain $\sum m_\nu=0.207^{+0.047}_{-0.052}~{\rm eV}$ for SdSDE+$\sum m_\nu$, reduced to $\sum m_\nu=0.162^{+0.055}_{-0.056}~{\rm eV}$ when $N_{\rm eff}$ is also varied. This arises from the positive correlation between $N_{\rm eff}$ and $\sum m_\nu$, together with the systematic preference of SdSDE for values of $N_{\rm eff}$ below the standard value. Furthermore, the best-fit $\chi^2$ comparison shows that $\Lambda$CDM with extended neutrino parameters is strongly preferred over the corresponding SdSDE extension. Overall, the positive neutrino mass preference induced by SdSDE may reflect parameter compensation rather than an improved global fit, a possibility that should be further tested with future high-precision observational data.

\end{abstract}

\maketitle

\section{Introduction}\label{sec1}

Measuring the absolute neutrino-mass scale has become a central goal in both particle physics and cosmology. Neutrino oscillation experiments have established that neutrinos have non-zero masses and can precisely measure the two independent mass squared differences among the three neutrino mass eigenstates~\cite{Super-Kamiokande:1998kpq,Super-Kamiokande:2000ywb,SNO:2001kpb,ParticleDataGroup:2014cgo,DayaBay:2022orm}; however, they cannot determine the absolute mass scale or the sign of the atmospheric mass splitting, leaving two possible mass orderings with different lower bounds on the total neutrino mass $\sum m_\nu$: the normal hierarchy (NH), corresponding to $\sum m_\nu \gtrsim 0.059~\mathrm{eV}$, and the inverted hierarchy, corresponding to $\sum m_\nu \gtrsim 0.10~\mathrm{eV}$~\cite{Esteban:2020cvm,ParticleDataGroup:2022pth,deSalas:2020pgw,Esteban:2024eli,Capozzi:2025wyn}. The first physics result from JUNO has recently improved the precision of oscillation-parameter measurements and mildly favors the NH, although no decisive determination has yet been reached~\cite{JUNO:2026first,Esteban:2026phq}.

Cosmological observations provide an independent approach for inferring $\sum m_\nu$ indirectly through the impact of massive neutrinos on the cosmic background evolution and the growth of structure. Neutrino free streaming suppresses structure growth below the free-streaming scale, leaving observable imprints on cosmic microwave background (CMB) anisotropies and lensing, galaxy clustering, and weak-lensing measurements; baryon acoustic oscillation (BAO) distance measurements further help constrain the background geometry and break parameter degeneracies~\cite{Lesgourgues:2006nd,Planck:2018vyg}. Cosmology therefore provides one of the most stringent constraints on $\sum m_\nu$. The recent Data Release~2 (DR2) of the Dark Energy Spectroscopic Instrument (DESI) has delivered high-precision BAO measurements, making cosmological constraints on neutrino mass particularly stringent. In the $\Lambda$CDM model, DESI BAO combined with CMB data gives a 95\% confidence-level upper limit of $\sum m_\nu<0.064~\mathrm{eV}$~\cite{DESI:2025zgx}, already very close to the lower limit required by oscillation experiments for the NH. This emerging tension between the cosmological upper bound and the particle-physics lower limit has become one of the most closely watched issues at the interface of cosmology and particle physics.

However, this neutrino-mass tension arises within the $\Lambda$CDM model. Although $\Lambda$CDM successfully describes a wide range of cosmological and astrophysical observations~\cite{Planck:2018vyg,eBOSS:2020yzd,Brout:2022vxf}, it still faces the ``fine-tuning'' and ``cosmic coincidence'' problems~\cite{Sahni:1999gb,Bean:2005ru}, as well as observational tensions that have emerged in recent years~\cite{Li:2013dha,Zhang:2014dxk,Zhao:2017urm,Guo:2018ans,Verde:2019ivm,DiValentino:2020zio,DiValentino:2020vvd,DiValentino:2019ffd,Vagnozzi:2019ezj,Vagnozzi:2021gjh,Vagnozzi:2021tjv,Vagnozzi:2023nrq,Gao:2021xnk,Jin:2021pcv,Riess:2021jrx,Schoneberg:2021qvd,Abdalla:2022yfr,DiValentino:2022fjm,Kamionkowski:2022pkx,Gao:2022ahg,Giare:2023xoc,Hu:2023jqc,Wright:2025xka,Cai:2026swf,Pantos:2026koc}, most notably the $H_0$ tension. If not explained by unaccounted-for systematic effects~\cite{Riess:2024ohe}, the $H_0$ tension may point to new physics beyond the standard $\Lambda$CDM model, with possible explanations including dynamical dark energy (DE)~\cite{Chevallier:2000qy,Kamenshchik:2001cp,Linder:2002et}, holographic DE~\cite{Li:2004rb,Huang:2004ai,Wang:2004nqa,Zhang:2005hs,Zhang:2005yz,Zhang:2009un,Zhang:2006qu,Fu:2011ab,Cui:2009ns,Ma:2007av,Zhang:2007sh,Wang:2016och}, interacting DE~\cite{Farrar:2003uw,Zhang:2004gc,Cai:2004dk,Zhang:2005rg,Zhang:2005kj,Zhang:2005rj,Wang:2006qw,Zhang:2012uu,Li:2014cee,Li:2015vla,Pan:2020zza,Yao:2022kub,Wang:2024vmw,Wang:2021kxc}, early DE~\cite{Poulin:2018cxd,Smith:2020rxx,Yin:2023srb,Bella:2026zuk,Ye:2020btb,Sabogal:2026qvy,Du:2026qtq}, modified gravity~\cite{Boisseau:2000pr,Clifton:2011jh,Cai:2015emx,Koyama:2015vza}, and others~\cite{Zhu:2023gmx,Wang:2025dzn,Du:2025odq,Lee:2022gzh,Lynch:2024hzh,Song:2025ddm,Prasanthan:2026boc}. The recent DESI results have posed a further challenge to the $\Lambda$CDM model. When combined with CMB and type Ia supernova (SN) data, the DESI BAO measurements show deviations from $\Lambda$CDM and a preference for dynamical DE~\cite{DESI:2025zgx}. The DESI BAO results have already motivated a broad range of cosmological analyses beyond $\Lambda$CDM~\cite{Colgain:2024xqj,Giare:2024smz,Li:2024qus,Li:2024qso,Jiang:2024xnu,Du:2024pai,Ye:2024ywg,Escamilla:2024ahl,Sabogal:2024yha,Li:2024bwr,Wang:2024dka,Wu:2025vfs,Huang:2025som,Li:2025owk,Li:2025ula,Ling:2025lmw,Pang:2025lvh,Ozulker:2025ehg,Cheng:2025lod,Pan:2025qwy,Liu:2025myr,Yang:2025ume,Chen:2025wwn,Abedin:2025dis,Li:2025eqh,Li:2025dwz,Li:2025cxn,Avila:2025sjz,Yang:2025uyv,vanderWesthuizen:2025vcb,FrancoAbellan:2025fkb,Paliathanasis:2025kmg,Li:2025htp,Hussain:2025uye,Samanta:2025oqz,Yang:2025oax,Paul:2025wix,Yao:2025kuz,Li:2025vuh,Song:2025bio,Du:2025csv,Zhang:2025dwu,Li:2025muv,Wang:2025djw,RoyChoudhury:2025iis,Luciano:2025ykr,Pedrotti:2025ccw,Pan:2025hbr,Afroz:2025iwo,Liu:2025evk,Li:2026hwq,Montefalcone:2026iga,Wang:2026rkx,Du:2026cly,Li:2026xaz,Wang:2026kbg,Piccoli:2026grg,Comini:2026nsj,Montero-Camacho:2026ohy,Song:2026kii,Jiang:2026cqh,Hashim:2026yoy,Wang:2025xvi,Giare:2024oil,Giare:2024gpk,Lee:2026yzs,Wang:2026wrk,Wang:2025znm,Wolf:2025jed,Yao:2025twv,Antusch:2026ldp,Xu:2026sbw,Giare:2026oti}; for recent reviews, see, e.g., Refs.~\cite{CosmoVerseNetwork:2025alb,Giare:2025pzu,Qiu:2025oop,Cai:2025mas,Anchordoqui:2026hys,Li:2026asg,Ren:2026jyw}. More importantly, the DESI analysis in dynamical DE models indicates that, once the DE equation of state (EoS) parameters $w_0$ and $w_a$ are allowed to vary, the inferred constraint on $\sum m_\nu$ can change appreciably relative to the minimal $\Lambda$CDM case, with the upper limit being substantially relaxed~\cite{DESI:2025zgx}.

This illustrates a broader model dependence of cosmological neutrino-mass inference, which has been systematically investigated in previous work~\cite{Zhang:2015uhk,Zhao:2016ecj,Zhang:2017rbg,Guo:2018gyo,Vagnozzi:2018jhn}. In particular, \citet{Zhang:2015uhk} performed systematic analyses across multiple DE models and showed that the inferred upper limit on $\sum m_\nu$ can vary significantly when the DE dynamics is changed. Related studies have also emphasized the degeneracy and parameter compensation between neutrino mass and DE parameters~\cite{Hannestad:2005gj,Wang:2016tsz,Yang:2017amu,RoyChoudhury:2018gay}. Recent studies further reinforce this picture from complementary perspectives. On one hand, several DESI-era extended cosmological analyses have reported a preference for a positive neutrino mass~\cite{DESI:2025ffm,Reeves:2025axp,RoyChoudhury:2025dhe,Du:2025xes,Sullivan:2026tas}. On the other hand, if the neutrino mass is generalized to an effective parameter, some data combinations may instead exhibit a preference for an effective negative neutrino mass~\cite{Elbers:2024sha,Elbers:2025vlz,Cozzumbo:2025ewt}. Such a ``negative mass'' should not be interpreted as a physical neutrino mass, but rather as a diagnostic signal of model-dependent parameter compensation~\cite{Green:2024xbb,Jhaveri:2025neg,Namikawa:2025doa,Kibris:2026cqq}. Taken together, the positive-mass preference and the negative-effective-mass diagnostic both indicate that cosmological neutrino mass constraints should be regarded as model-dependent inferences whose robustness must be tested under different DE assumptions and data combinations~\cite{Mainini:2010ng,Li:2012spm,Zhang:2014ifa,Li:2015poa,Zhang:2017rbg,Zhang:2014nta,Zhang:2015uhk,Zhao:2016ecj,Feng:2017usu,Feng:2017nss,Guo:2018gyo,Vagnozzi:2018jhn,DiazRivero:2019ukx,Feng:2019jqa,Liu:2020vgn,Yao:2022jrw,Chudaykin:2022rnl,Sharma:2022ifr,Li:2023gtu,Jiang:2024viw,Barua:2025adv,Zhou:2025nkb,Du:2025iow,Feng:2025mlo,Yang:2026yaq,Ladeira:2026jne,Feng:2026pzs,Li:2026ldf,Gil-Ocaranza:2026lpq,Hou:2026phk,Yang:2026kxp,Wang:2026gvg}.

Among attempts to go beyond a purely phenomenological description of DE, black-hole-inspired DE scenarios provide an intriguing possibility. Black holes are strong gravity systems, and the classical singularity problem suggests that general relativity may require an effective description beyond its classical regime in extreme gravitational environments. In this context, cosmologically coupled black-hole models have explored whether black-hole populations may contribute an effective DE component on cosmological scales~\cite{Alexander:2019ctv,Croker:2019kje,Croker:2019mup,Croker:2020plg,Croker:2021duf,Farrah:2023opk,Croker:2024jfg}. Although the theoretical interpretation and observational viability of these scenarios remain under active discussion, they provide a route to the physical origin of DE that differs from conventional scalar-field models or phenomenological EoS parameterizations. We focus on the Schwarzschild--de Sitter black-hole dark energy (SdSDE) model proposed by~\citet{Hayashi:2025frr}. The model is motivated by the possibility that local de Sitter-like contributions associated with non-singular black holes may be negligible compared with the rest mass of an individual black hole, while the cumulative contribution from the cosmic black-hole population can be phenomenologically described as an effective DE density. The DE evolution in SdSDE is tied to the redshift evolution of the cosmic black-hole mass density, leading to a fixed effective EoS $w_{\rm DE}(z)$. Therefore, SdSDE provides a concrete black-hole-inspired DE template at the background level.

In this work, we examine the cosmological inference of $\sum m_\nu$ within the SdSDE framework, using CMB data from Planck and the Atacama Cosmology Telescope (ACT) DR6, DESI DR2 BAO measurements, and SN data from DES-Dovekie and PantheonPlus. We further investigate the impact of allowing $N_{\rm eff}$ to vary on $\sum m_\nu$ constraints, and assess whether SdSDE leads to distinct inferences for key neutrino parameters compared with the corresponding $\Lambda$CDM extensions.

This paper is organized as follows. In Sec.~\ref{sec2}, we briefly introduce the SdSDE model and the cosmological data used in the analysis. In Sec.~\ref{sec3}, we report the constraint results and provide relevant discussions. The conclusion is given in Sec.~\ref{sec4}.

\section{Methodology}\label{sec2}

\subsection{Schwarzschild--de Sitter black-hole dark energy model}

In this work, we adopt the SdSDE model proposed in Ref.~\cite{Hayashi:2025frr}. At the background level, the key input of the SdSDE model is the redshift evolution of the cosmic black-hole mass density. Following Ref.~\cite{Hayashi:2025frr}, the effective DE density in this model can be written as
\begin{equation}
\rho_{\rm DE}(z)=\rho_{\rm DE,0}\exp[\mathcal{F}(z)],
\end{equation}
where $\rho_{\rm DE,0}$ is the present-day effective DE density in the SdSDE model, and $\mathcal{F}(z)$ is determined by a polynomial fit to the black-hole mass-density evolution. The original SdSDE analysis adopts the cubic form
\begin{equation}
\mathcal{F}(z)=a z^3+b z^2+c z.
\label{eq:fz}
\end{equation}

The coefficients are fixed to
\begin{equation}
a=0.00658,\qquad b=-0.104,\qquad c=-0.348.
\end{equation}

These values are taken from the polynomial fit to the cosmic black-hole mass density adopted in Ref.~\cite{Hayashi:2025frr}, based on the black-hole mass-function modeling of Ref.~\cite{Sicilia:2021gtu}. The constant term of the fit is absorbed into the present-day normalization $\rho_{\rm DE,0}$, so that only the redshift-dependent shape enters our analysis. Consequently, the SdSDE background is treated as a prescribed phenomenological model, rather than being reconstructed from cosmological data. Since this description is tied to the astrophysical black-hole population, it should be applied only after black-hole formation becomes relevant and should not be extrapolated to arbitrarily high redshifts. In our numerical implementation, we therefore introduce a smooth cutoff around $z_{\rm cut}\simeq10$, such that the model continuously approaches $\Lambda$CDM at higher redshifts.

The corresponding effective EoS follows from the usual DE continuity equation,
\begin{equation}
\dot{\rho}_{\rm DE}+3H(1+w_{\rm DE})\rho_{\rm DE}=0,
\end{equation}
where $H$ is the Hubble expansion rate and the overdot denotes the derivative with respect to cosmic time.

It can be written as
\begin{equation}
w_{\rm DE}(z)=-1+\frac{1+z}{3}\frac{\mathrm{d}\ln\rho_{\rm DE}(z)}{\mathrm{d}z}
=-1+\frac{1+z}{3}\frac{\mathrm{d}\mathcal{F}(z)}{\mathrm{d}z}.
\end{equation}

From Eq.~(\ref{eq:fz}), this gives
\begin{equation}
w_{\rm DE}(z)
=-1+\frac{1+z}{3}\left(3a z^2+2b z+c\right).
\end{equation}

Thus, once the coefficients $a$, $b$, and $c$ are fixed, the SdSDE model predicts a specified redshift dependence of $w_{\rm DE}(z)$.

\subsection{Massive neutrinos and structure formation}

We assume a standard thermal history with three active neutrino species. In the early Universe, neutrinos behave as a relativistic component and contribute to the radiation density. The relativistic neutrino energy density is conventionally parameterized as
\begin{equation}
\rho_\nu = N_{\rm eff}\frac{7}{8}\left(\frac{4}{11}\right)^{4/3}\rho_\gamma,
\end{equation}
where $\rho_\gamma$ is the photon energy density and $N_{\rm eff}$ is the effective number of relativistic species, whose standard value is $N_{\rm eff}=3.044$. Thus, varying $N_{\rm eff}$ changes the early radiation density, the expansion rate, and the sound horizon.

In the late Universe, after becoming non-relativistic, massive neutrinos contribute to the matter density, with the present-day density parameter given by
\begin{equation}
\Omega_\nu h^2=\frac{\sum m_\nu}{93.14\,{\rm eV}},
\end{equation}
where $h$ is the dimensionless Hubble parameter, defined by $h=H_0/(100~{\rm km\,s^{-1}\,Mpc^{-1}})$.

Massive neutrinos affect structure formation mainly through free streaming. Their large thermal velocities prevent them from clustering efficiently below the free-streaming scale. A commonly used estimate of the corresponding free-streaming wavenumber is
\begin{equation}
k_{\mathrm{fs}}(z) \simeq 0.018\, \sqrt{\frac{\Omega_{\rm m} (1+z)}{0.3}} \left( \frac{\sum m_\nu}{1.0\,\mathrm{eV}} \right) h\, \mathrm{Mpc}^{-1},
\end{equation}
where $\Omega_{\rm m}$ is the present-day total matter density parameter.

On scales much smaller than the free-streaming scale, neutrino perturbations are suppressed, reducing the total matter power spectrum $P(k)$ relative to a massless-neutrino cosmology. In the small neutrino-fraction limit, the approximate suppression can be written as
\begin{equation}
\frac{\Delta P(k)}{P(k)} \approx -8\frac{\Omega_\nu}{\Omega_{\rm m}}.
\end{equation}

This suppression lowers the amplitude of matter clustering and hence affects observables such as $\sigma_8$, $S_8$, and CMB lensing. The variance of the matter density field smoothed on a comoving scale $R$ is
\begin{equation}
\sigma^2(R)=\frac{1}{2\pi^2}\int k^2P(k)\left[\frac{3j_1(kR)}{kR}\right]^2 \mathrm{d}k,
\end{equation}
where $j_1(x)=(\sin x-x\cos x)/x^2$. Evaluating this quantity at $R=8\,h^{-1}{\rm Mpc}$ gives $\sigma_8$ and the commonly quoted parameter $S_8=\sigma_8\sqrt{\Omega_{\rm m}/0.3}$.

\begin{table}[t]
\caption{Flat priors on the main cosmological parameters varied in this work.}
\label{tab:priors}
\begin{center}
\renewcommand{\arraystretch}{1.35}
\begin{tabular}{@{\hspace{0.5cm}} c @{\hspace{1.0cm}} c @{\hspace{0.8cm}}}
\bottomrule[0.8pt]
\bottomrule[0.8pt]
Parameter & Prior \\
\bottomrule[0.8pt]
$\Omega_{\rm b}h^2$                   & $\mathcal{U}[0.005,\,0.1]$ \\
$\Omega_{\rm c}h^2$                   & $\mathcal{U}[0.001,\,0.99]$ \\
$H_0~[\mathrm{km\,s^{-1}\,Mpc^{-1}}]$ & $\mathcal{U}[20,\,100]$ \\
$\tau$                                & $\mathcal{U}[0.01,\,0.8]$ \\
$\ln(10^{10}A_{\rm s})$               & $\mathcal{U}[1.61,\,3.91]$ \\
$n_{\rm s}$                           & $\mathcal{U}[0.8,\,1.2]$ \\
$\sum m_\nu~[\mathrm{eV}]$            & $\mathcal{U}[0,\,5]$ \\
$N_{\rm eff}$                         & $\mathcal{U}[0,\,7]$ \\
\bottomrule[0.8pt]
\end{tabular}
\end{center}
\end{table}

\begin{table*}[t]
\renewcommand\arraystretch{1.55}
\setlength{\tabcolsep}{2.4pt}
\centering
\footnotesize
\caption{The $1\sigma$ confidence regions of cosmological parameters for the SdSDE, SdSDE+$\sum m_\nu$, SdSDE+$N_{\rm eff}$, and SdSDE+$\sum m_\nu+N_{\rm eff}$ models, obtained from the CMB+DESI+DES-Dovekie data.}
\label{tab:sdsde_hierarchy}
\begin{tabular*}{\textwidth}{@{\extracolsep{\fill}}lccccc@{}}
\toprule[0.8pt]
\toprule[0.8pt]
\textbf{Model} & \textbf{$H_0$ [$\mathrm{km\,s^{-1}\,Mpc^{-1}}$]} & \textbf{$S_8$} & \textbf{$\Omega_{\rm m}$} & \textbf{$\sum m_{\nu}$ [$\mathrm{eV}$]} & \textbf{$N_{\rm eff}$} \\
\midrule[0.8pt]
SdSDE & $71.85^{+0.33}_{-0.32}$ & $0.8391\pm0.0071$ & $0.2814^{+0.0034}_{-0.0036}$ & $-$ & $-$ \\
SdSDE+$\sum m_\nu$ & $71.10^{+0.40}_{-0.39}$ & $0.8153^{+0.0110}_{-0.0112}$ & $0.2874^{+0.0039}_{-0.0041}$ & $0.207^{+0.047}_{-0.052}$ & $-$ \\
SdSDE+$N_{\rm eff}$ & $69.05^{+0.94}_{-0.92}$ & $0.8262\pm0.0082$ & $0.2869\pm0.0040$ & $-$ & $2.57\pm0.15$ \\
SdSDE+$\sum m_\nu+N_{\rm eff}$ & $69.46^{+0.93}_{-1.01}$ & $0.8142^{+0.0104}_{-0.0106}$ & $0.2889^{+0.0040}_{-0.0041}$ & $0.162^{+0.055}_{-0.056}$ & $2.72\pm0.18$ \\
\bottomrule[0.8pt]
\end{tabular*}
\end{table*}

\subsection{Cosmological datasets}

In this work, the theoretical predictions are computed using a modified version of the Boltzmann solver {\tt CAMB}\footnote{\url{https://github.com/cmbant/CAMB}.}~\cite{Lewis:1999bs,Howlett:2012mh}. Bayesian inference is performed with the cosmological inference code {\tt Cobaya}\footnote{\url{https://github.com/CobayaSampler/cobaya}.}~\cite{Torrado:2020dgo}, using Markov chain Monte Carlo (MCMC) sampling. The convergence of the MCMC chains is assessed with the Gelman--Rubin criterion, and we require $R-1<0.02$ for the chains used in the final analysis. The posterior distributions are analyzed with the public package {\tt GetDist}\footnote{\url{https://github.com/cmbant/getdist}.}~\cite{Lewis:2019xzd}.

For both $\Lambda$CDM and SdSDE, the baseline cosmological parameter set varied in the MCMC analysis is $\boldsymbol{\theta}_{\rm base} = \{ \Omega_{\rm b}h^2, \Omega_{\rm c}h^2, H_0, \tau, \ln(10^{10}A_{\rm s}), n_{\rm s} \}$. For the neutrino parameters, we additionally vary $\sum m_\nu$ and $N_{\rm eff}$, depending on the model under consideration, and adopt a degenerate-mass treatment for massive neutrinos. The prior ranges used in this work are summarized in Table~\ref{tab:priors}.

We utilize the following cosmological observational datasets:

\begin{itemize}

\item \textbf{CMB}. This analysis incorporates measurements of the CMB temperature anisotropy, polarization power spectra, their cross-spectra, and the ACT lensing power spectrum. The CMB likelihoods comprise four components: (i) The small-scale ($\ell>30$) temperature and polarization anisotropy spectra, $C_{\ell}^{TT}$, $C_{\ell}^{TE}$, and $C_{\ell}^{EE}$, are taken from the Planck PR4 \texttt{CamSpec} likelihood~\cite{Planck:2018vyg,Efstathiou:2019mdh,Rosenberg:2022sdy}. (ii) The large-scale ($2\leq \ell \leq 30$) temperature anisotropy power spectrum $C_{\ell}^{TT}$ is described by the Planck \texttt{Commander} likelihood~\cite{Planck:2018vyg,Planck:2019nip}. (iii) The large-scale ($2\leq \ell \leq 30$) E-mode polarization spectrum $C_{\ell}^{EE}$ is described by the Planck \texttt{SimAll} likelihood~\cite{Planck:2018vyg,Planck:2019nip}. (iv) The CMB lensing likelihood is based on the latest high-precision reconstruction from NPIPE PR4 Planck data and ACT Data Release 6~\cite{ACT:2023dou}.

\item \textbf{DESI}. The DESI DR2 BAO measurements are based on tracers including the bright galaxy sample, luminous red galaxies, emission-line galaxies, quasars, and the Lyman-$\alpha$ forest, as summarized in Table IV of Ref.~\cite{DESI:2025zgx}. 

\item \textbf{DES-Dovekie}. The DES-Dovekie dataset is a reconstructed SN dataset obtained from a recalibration and reanalysis of the DESY5 SN sample. It contains 1623 likely type Ia supernovae (SNe) with $P_{\rm SNe}>0.5$ and 197 low-redshift SNe, giving a total of 1820 SNe~\cite{DES:2025sig}.

\item \textbf{PantheonPlus}. The PantheonPlus sample comprises 1550 spectroscopically confirmed SNe from 18 different surveys, covering the redshift range $0.01<z<2.26$~\cite{Brout:2022vxf}. 
\end{itemize}

\section{Results and Discussion}\label{sec3}

\begin{figure}[t]
\centering
\includegraphics[width=\columnwidth]{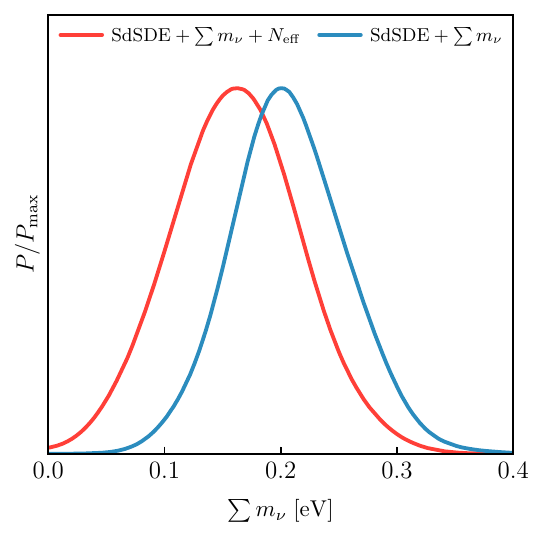}
\caption{
One-dimensional marginalized posterior distributions of $\sum m_\nu$ for SdSDE+$\sum m_\nu$ and SdSDE+$\sum m_\nu+N_{\rm eff}$ using the CMB+DESI+DES-Dovekie data.
}
\label{fig1}
\end{figure}

\begin{figure*}[t]
\centering
\includegraphics[width=\textwidth]{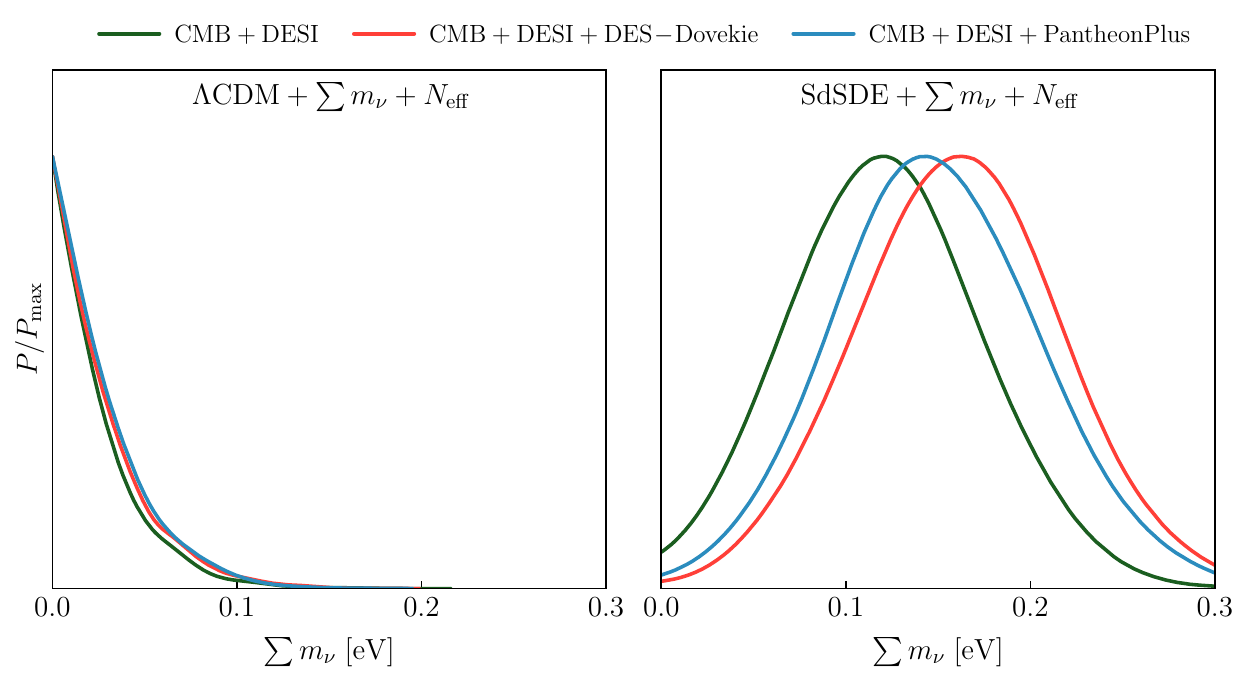}
\caption{One-dimensional marginalized posterior distributions of $\sum m_\nu$ in $\Lambda$CDM+$\sum m_\nu+N_{\rm eff}$ and SdSDE+$\sum m_\nu+N_{\rm eff}$ for the CMB+DESI, CMB+DESI+DES-Dovekie, and CMB+DESI+PantheonPlus data.}
\label{fig2}
\end{figure*}

\begin{figure*}[t]
\centering
\includegraphics[width=0.9\textwidth]{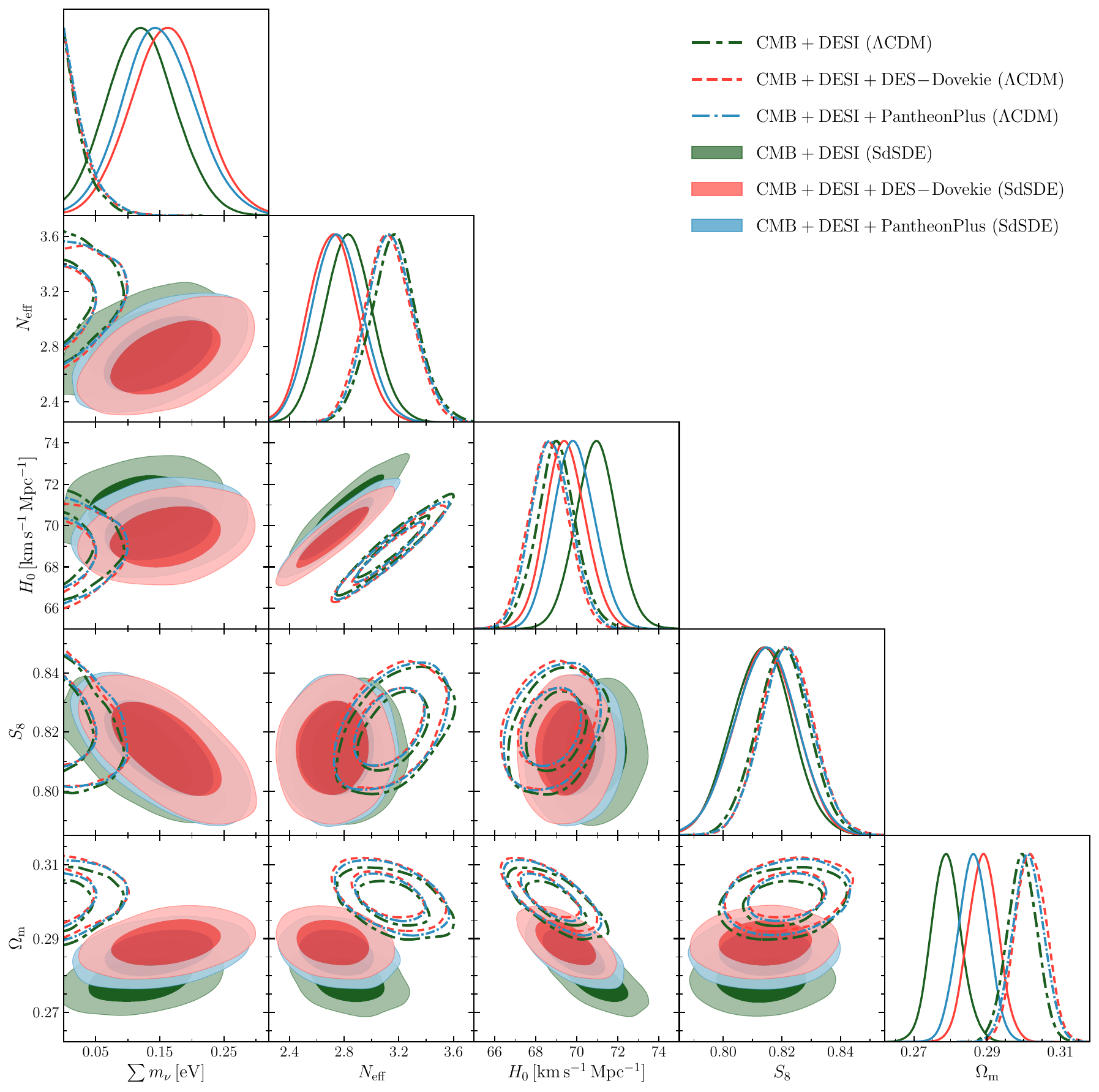}
\caption{Constraints on the cosmological parameters using the CMB+DESI, CMB+DESI+DES-Dovekie, and CMB+DESI+PantheonPlus data in the $\Lambda$CDM+$\sum m_\nu+N_{\rm eff}$ (line contours) and SdSDE+$\sum m_\nu+N_{\rm eff}$ (filled contours) models.}
\label{fig3}
\end{figure*}

\begin{figure*}[t]
\centering
\includegraphics[width=\textwidth]{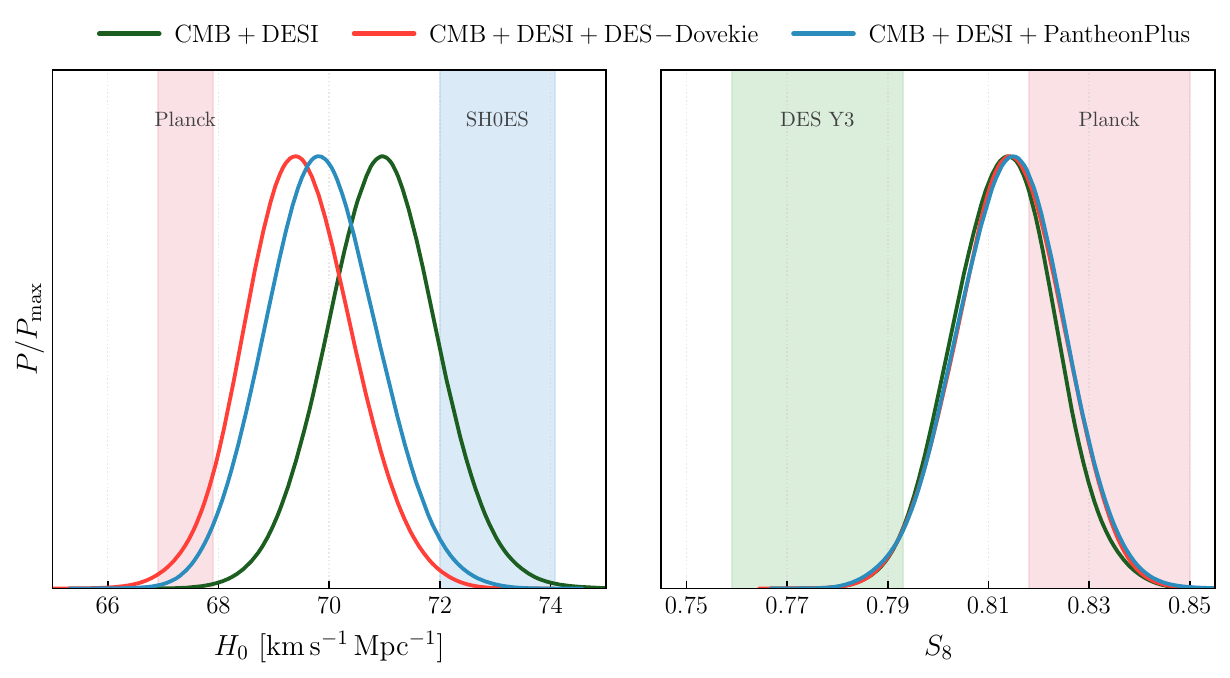}
\caption{One-dimensional marginalized posterior distributions of $H_0$ and $S_8$ for SdSDE+$\sum m_\nu+N_{\rm eff}$ using the CMB+DESI, CMB+DESI+DES-Dovekie, and CMB+DESI+PantheonPlus data. The shaded bands indicate the corresponding Planck~\cite{Planck:2018vyg}, SH0ES~\cite{Riess:2021jrx}, and DES Y3~\cite{DES:2021wwk} reference constraints.}
\label{fig4}
\end{figure*}

\begin{figure}[t]
\centering
\includegraphics[width=\columnwidth]{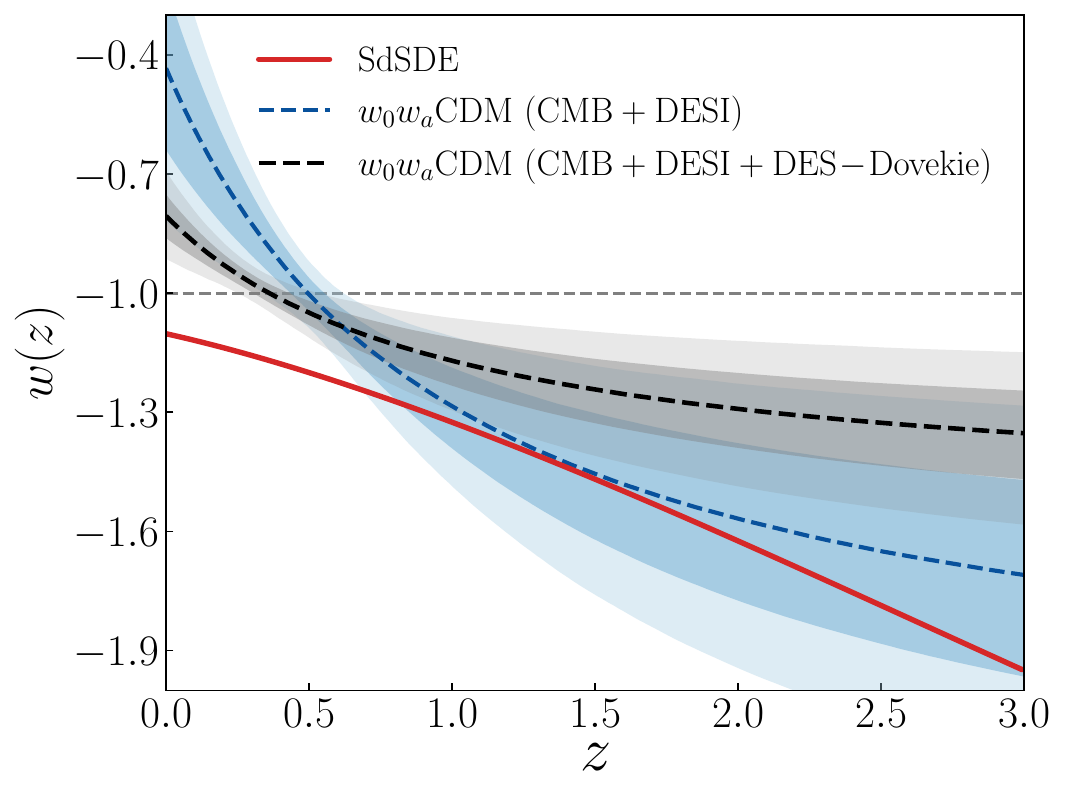}
\caption{Comparison of the DE EoS evolution $w(z)$ between the fixed SdSDE prediction and the phenomenological $w_0w_a$CDM reconstruction using the CMB+DESI and CMB+DESI+DES-Dovekie data. The shaded regions indicate the $1\sigma$ and $2\sigma$ confidence intervals.}
\label{fig5}
\end{figure}

In this section, we present the main cosmological constraint results, as summarized in Tables~\ref{tab:sdsde_hierarchy} and \ref{tab:joint_neutrino_constraints} and Figs.~\ref{fig1}--\ref{fig4}. We also compare the EoS of SdSDE evolution with the $w_0w_a$CDM reconstruction in Fig.~\ref{fig5}. Finally, the goodness-of-fit results are summarized in Table~\ref{tab:chi2}.

\begin{table*}[t]
\renewcommand\arraystretch{1.55}
\setlength{\tabcolsep}{2.6pt}
\centering
\footnotesize
\caption{The $1\sigma$ confidence regions (or $2\sigma$ upper limits) of cosmological parameters obtained by the CMB, DESI, DES-Dovekie, and PantheonPlus data for the $\Lambda$CDM+$\sum m_\nu$+$N_{\rm eff}$ and SdSDE+$\sum m_\nu$+$N_{\rm eff}$ models. For the parameter $\sum m_\nu$, central values cannot be determined in most cases, and we provide the $2\sigma$ upper limits.}
\label{tab:joint_neutrino_constraints}
\begin{tabular*}{\textwidth}{@{\extracolsep{\fill}}lccccc@{}}
\toprule[0.8pt]
\toprule[0.8pt]
\textbf{Data} & \textbf{$H_0$ [$\mathrm{km\,s^{-1}\,Mpc^{-1}}$]} & \textbf{$S_8$} & \textbf{$\Omega_{\rm m}$} & \textbf{$\sum m_{\nu}$ [$\mathrm{eV}$]} & \textbf{$N_{\rm eff}$} \\
\midrule[0.8pt]
\multicolumn{6}{l}{\textbf{$\Lambda$CDM+$\sum m_\nu+N_{\rm eff}$}} \\
CMB+DESI & $69.01^{+0.93}_{-0.95}$ & $0.8203\pm0.0088$ & $0.2996\pm0.0041$ & $<0.073$ & $3.16^{+0.16}_{-0.17}$ \\
CMB+DESI+DES-Dovekie & $68.63^{+0.97}_{-0.98}$ & $0.8218^{+0.0087}_{-0.0089}$ & $0.3020^{+0.0042}_{-0.0041}$ & $<0.080$ & $3.12\pm0.17$ \\
CMB+DESI+PantheonPlus & $68.76^{+0.98}_{-0.97}$ & $0.8216^{+0.0088}_{-0.0087}$ & $0.3012\pm0.0042$ & $<0.079$ & $3.14^{+0.16}_{-0.18}$ \\
\midrule[0.4pt]
\multicolumn{6}{l}{\textbf{SdSDE+$\sum m_\nu+N_{\rm eff}$}} \\
CMB+DESI & $70.95^{+0.98}_{-0.99}$ & $0.8132^{+0.0099}_{-0.0102}$ & $0.2789\pm0.0040$ & $0.122^{+0.049}_{-0.056}$ & $2.83\pm0.17$ \\
CMB+DESI+DES-Dovekie & $69.46^{+0.93}_{-1.01}$ & $0.8142^{+0.0104}_{-0.0106}$ & $0.2889^{+0.0040}_{-0.0041}$ & $0.162^{+0.055}_{-0.056}$ & $2.72\pm0.18$ \\
CMB+DESI+PantheonPlus & $69.87\pm1.01$ & $0.8144^{+0.0104}_{-0.0106}$ & $0.2862\pm0.0041$ & $0.149^{+0.054}_{-0.058}$ & $2.75\pm0.18$ \\
\bottomrule[0.8pt]
\end{tabular*}
\end{table*}

We first report the constraints on the baseline SdSDE model and on its extensions with free $\sum m_\nu$, free $N_{\rm eff}$, and both parameters varied simultaneously, using the representative CMB+DESI+DES-Dovekie data, as shown in Table~\ref{tab:sdsde_hierarchy}. The baseline SdSDE model gives $H_0=71.85^{+0.33}_{-0.32}~\mathrm{km\,s^{-1}\,Mpc^{-1}}$ and $S_8=0.8391\pm0.0071$, suggesting that the SdSDE background tends to favor a relatively high late-time expansion rate and a large clustering amplitude. When only $\sum m_\nu$ is allowed to vary, we obtain $\sum m_\nu=0.207^{+0.047}_{-0.052}~{\rm eV}$, corresponding to a preference of about $4.0\sigma$ for a positive neutrino mass, with the posterior in Fig.~\ref{fig1} peaking away from the lower prior boundary. Meanwhile, $S_8$ decreases to $0.8153^{+0.0110}_{-0.0112}$, consistent with the free-streaming suppression of structure growth by massive neutrinos. When only $N_{\rm eff}$ is varied, the constraint becomes $N_{\rm eff}=2.57\pm 0.15$, below the standard value, with $H_0$ reduced to $69.05^{+0.94}_{-0.92}~\mathrm{km\,s^{-1}\,Mpc^{-1}}$. Thus, the preferred direction is not toward extra radiation, but toward a lower early-time radiation density. When both $\sum m_\nu$ and $N_{\rm eff}$ are varied, we find $\sum m_\nu=0.162^{+0.055}_{-0.056}~{\rm eV}$ and $N_{\rm eff}=2.72\pm0.18$. The corresponding descriptive significance of the neutrino-mass preference is reduced to about $2.9\sigma$. Therefore, allowing $N_{\rm eff}$ to vary weakens the preference for a positive $\sum m_\nu$, but does not remove it.

We next examine whether this behavior persists across different data combinations and compare it with the corresponding results in $\Lambda$CDM+$\sum m_\nu+N_{\rm eff}$. As shown in Table~\ref{tab:joint_neutrino_constraints} and Fig.~\ref{fig2}, in the $\Lambda$CDM+$\sum m_\nu+N_{\rm eff}$ case, all three data combinations yield only upper limits on $\sum m_\nu$. By contrast, SdSDE+$\sum m_\nu+N_{\rm eff}$ gives $\sum m_\nu=0.122^{+0.049}_{-0.056}~{\rm eV}$ for CMB+DESI, corresponding to a preference for a positive value of $\sum m_\nu$ at about $2.2\sigma$; $\sum m_\nu=0.162^{+0.055}_{-0.056}~{\rm eV}$ for CMB+DESI+DES-Dovekie, corresponding to about $2.9\sigma$; and $\sum m_\nu=0.149^{+0.054}_{-0.058}~{\rm eV}$ for CMB+DESI+PantheonPlus, corresponding to about $2.6\sigma$.

\begin{table*}[t]
\renewcommand\arraystretch{1.55}
\setlength{\tabcolsep}{2.6pt}
\centering
\footnotesize
\caption{Summary of the goodness of fit for SdSDE+$\sum m_\nu+N_{\rm eff}$ relative to $\Lambda$CDM+$\sum m_\nu+N_{\rm eff}$. For each data combination, we report the best-fit $\chi^2$ contributions from CMB, BAO, and SN, the total best-fit $\chi^2$, and the corresponding $\Delta\chi^2$.}
\label{tab:chi2}
\begin{tabular*}{\textwidth}{@{\extracolsep{\fill}}llccccc@{}}
\toprule[0.8pt]
\toprule[0.8pt]
\textbf{Data} & \textbf{Model} & \textbf{$\chi^2_{\rm CMB}$} & \textbf{$\chi^2_{\rm BAO}$} & \textbf{$\chi^2_{\rm SN}$} & \textbf{$\chi^2_{\rm tot}$} & \textbf{$\Delta\chi^2$} \\
\midrule[0.8pt]
\multirow{2}{*}{CMB+DESI} & $\Lambda$CDM+$\sum m_\nu+N_{\rm eff}$ & $10985.53$ & $10.53$ & $-$ & $10996.06$ & $0.00$ \\
& SdSDE+$\sum m_\nu+N_{\rm eff}$ & $10982.81$ & $31.90$ & $-$ & $11014.71$ & $18.65$ \\
\midrule[0.4pt]
\multirow{2}{*}{CMB+DESI+DES-Dovekie} & $\Lambda$CDM+$\sum m_\nu+N_{\rm eff}$ & $10984.79$ & $10.59$ & $1636.20$ & $12631.58$ & $0.00$ \\
& SdSDE+$\sum m_\nu+N_{\rm eff}$ & $10985.37$ & $35.18$ & $1677.28$ & $12697.84$ & $66.26$ \\
\midrule[0.4pt]
\multirow{2}{*}{CMB+DESI+PantheonPlus} & $\Lambda$CDM+$\sum m_\nu+N_{\rm eff}$ & $10984.60$ & $11.49$ & $1405.93$ & $12402.02$ & $0.00$ \\
& SdSDE+$\sum m_\nu+N_{\rm eff}$ & $10984.99$ & $35.80$ & $1437.36$ & $12458.16$ & $56.14$ \\
\bottomrule[0.8pt]
\end{tabular*}
\end{table*}

To further understand the interplay between neutrino parameters and other cosmological parameters, Fig.~\ref{fig3} presents the constraints on the key parameters. Relative to the corresponding $\Lambda$CDM extension, SdSDE exhibits a consistent compensation direction across all three data combinations: $\sum m_\nu$ is driven toward a larger positive central value, $N_{\rm eff}$ is pushed below its standard value, while $H_0$ is maintained at a relatively high level, and $S_8$ is somewhat reduced. Under the CMB and BAO distance calibration, a larger neutrino mass typically shifts the geometric constraints toward a lower $H_0$, partially offsetting the tendency of the SdSDE background itself to raise $H_0$. As shown in the left panel of Fig.~\ref{fig4}, the most favorable case is obtained for CMB+DESI, for which SdSDE+$\sum m_\nu+N_{\rm eff}$ gives $H_0=70.95^{+0.98}_{-0.99}~{\rm km\,s^{-1}\,Mpc^{-1}}$, reducing the tension with the SH0ES measurement to about $1.5\sigma$. Meanwhile, a larger $\sum m_\nu$ can suppress late-time structure growth through neutrino free-streaming, thereby lowering $S_8$ relative to the baseline SdSDE model. As shown in the right panel of Fig.~\ref{fig4}, this shift moves the result mildly toward the lower-$S_8$ direction preferred by DES Y3, although the posterior still remains closer to the Planck reference constraint. On the other hand, the positive correlation between $\sum m_\nu$ and $N_{\rm eff}$ indicates that when $N_{\rm eff}$ is allowed to shift toward lower values, the preference for a larger $\sum m_\nu$ in SdSDE is correspondingly weakened. This explains why varying only $\sum m_\nu$ yields a higher central value for the neutrino mass, whereas simultaneously freeing $N_{\rm eff}$ reduces this value. Notably, SdSDE systematically pushes $N_{\rm eff}$ below its standard value. A lower $N_{\rm eff}$ corresponds to a lower neutrino number density or a colder neutrino background. One possible physical interpretation is that incomplete neutrino thermalization in the early Universe reduces the neutrino energy density relative to the photon energy density, thereby lowering the value of $N_{\rm eff}$ inferred from cosmological observations.

We further compare the fixed SdSDE EoS template with the phenomenological $w_0w_a$CDM reconstruction, as shown in Fig.~\ref{fig5}. Compared with the flexible $w_0w_a$CDM reconstruction, the SdSDE EoS is fixed by the model construction and tends to realize a phantom-like evolution of DE EoS. Such a background evolution modifies the late-time cosmic expansion by enhancing the effective DE density at low redshifts, thereby shifting the geometric constraints toward a region that can accommodate a higher value of $H_0$. At the same time, the resulting change in the background evolution can be partially compensated by a larger neutrino mass, leading to a positive shift in the inferred value of $\sum m_\nu$.

Finally, we compare the goodness of fit between SdSDE+$\sum m_\nu+N_{\rm eff}$ and $\Lambda$CDM+$\sum m_\nu+N_{\rm eff}$. Since the two models have the same number of free parameters, we directly compare their best-fit $\chi^2$ values, as shown in Table~\ref{tab:chi2}. For the CMB+DESI data, SdSDE+$\sum m_\nu+N_{\rm eff}$ gives $\Delta\chi^2=18.65$ relative to $\Lambda$CDM+$\sum m_\nu+N_{\rm eff}$. After further including DES-Dovekie and PantheonPlus, this difference increases to $\Delta\chi^2=66.26$ and $\Delta\chi^2=56.14$, respectively. These results show that the overall fit of SdSDE is worse than that of the corresponding $\Lambda$CDM extension. The $\chi^2$ decomposition further shows that the difference in the CMB contribution is relatively small, while the main problem comes from the BAO and SN parts, for which SdSDE gives significantly larger $\chi^2$ values than the corresponding $\Lambda$CDM extension. This indicates that the rigidity of the SdSDE EoS makes it difficult to effectively mimic the late-time dynamical behavior preferred by current BAO and SN data, which is also consistent with the DE EoS comparison shown in Fig.~\ref{fig5}.

\section{Conclusions}\label{sec4}

In this work, we use CMB data from Planck and ACT, DESI DR2 BAO measurements, and SN data from DES-Dovekie and PantheonPlus to investigate cosmological constraints on the neutrino parameters $\sum m_\nu$ and $N_{\rm eff}$ within the SdSDE model.

We find that a robust preference for a positive neutrino mass can be obtained within the SdSDE framework. For the representative CMB+DESI+DES-Dovekie data, SdSDE+$\sum m_\nu$ gives $\sum m_\nu=0.207^{+0.047}_{-0.052}~{\rm eV}$, corresponding to a positive-mass preference at about the $4.0\sigma$ level. When both $\sum m_\nu$ and $N_{\rm eff}$ are allowed to vary, we obtain $\sum m_\nu=0.162^{+0.055}_{-0.056}~{\rm eV}$. Meanwhile, SdSDE systematically prefers values of $N_{\rm eff}$ below the standard prediction, with typical constraints around $N_{\rm eff}\simeq2.7$--$2.8$. Since $N_{\rm eff}$ is positively correlated with $\sum m_\nu$, a lower $N_{\rm eff}$ weakens the preference for a positive neutrino mass, although the preference remains at about the $2.9\sigma$ level. This result is associated with the phantom-like evolution of the EoS in SdSDE; such an evolution can induce a positive central value of $\sum m_\nu$ through parameter compensation.

However, the best-fit $\chi^2$ comparison shows that the current form of SdSDE does not improve the overall cosmological fit. Relative to the corresponding extended $\Lambda$CDM model, SdSDE gives a significantly larger best-fit $\chi^2$, with the degradation mainly coming from late-time distance measurements such as BAO and SN data, rather than from the CMB contribution. This may be because the fixed SdSDE EoS cannot sufficiently reproduce the quintom-like evolution crossing $w=-1$ that is preferred by current BAO and SN data. Looking ahead, making black-hole-inspired DE models more competitive may require introducing effective degrees of freedom that allow phantom-divide crossing while preserving the underlying black-hole motivation, together with further tests using more comprehensive high-precision data.

\section*{ACKNOWLEDGMENTS}
We thank Shintaro K. Hayashi and Hui Li for helpful discussions. This work was supported by the National Natural Science Foundation of China (Grants Nos. 12533001, 12575049, and 12473001), the National SKA Program of China (Grants Nos. 2022SKA0110200 and 2022SKA0110203), the China Manned Space Program (Grant No. CMS-CSST-2025-A02), and the National 111 Project (Grant No. B16009).

\bibliography{main}

\end{document}